# Functional Mechanism: Regression Analysis under Differential Privacy

Jun Zhang[1]    Zhenjie Zhang[2]    Xiaokui Xiao[1]    Yin Yang[2]    Marianne Winslett[2,3]

[1]School of Computer Engineering
Nanyang Technological University
{jzhang027,xkxiao}@ntu.edu.sg

[2]Advanced Digital Sciences Center
Illinois at Singapore Pte. Ltd.
{zhenjie,yin.yang}@adsc.com.sg

[3]Department of Computer Science
University of Illinois at Urbana-Champaign
winslett@illinois.edu

## ABSTRACT

$\epsilon$-*differential privacy* is the state-of-the-art model for releasing sensitive information while protecting privacy. Numerous methods have been proposed to enforce $\epsilon$-differential privacy in various analytical tasks, e.g., *regression analysis*. Existing solutions for regression analysis, however, are either limited to non-standard types of regression or unable to produce accurate regression results. Motivated by this, we propose the *Functional Mechanism*, a differentially private method designed for a large class of optimization-based analyses. The main idea is to enforce $\epsilon$-differential privacy by perturbing the *objective function* of the optimization problem, rather than its results. As case studies, we apply the functional mechanism to address two most widely used regression models, namely, *linear regression* and *logistic regression*. Both theoretical analysis and thorough experimental evaluations show that the functional mechanism is highly effective and efficient, and it significantly outperforms existing solutions.

## 1. INTRODUCTION

Releasing sensitive data while protecting privacy has been a subject of active research for the past few decades. One state-of-the-art approach to the problem is $\epsilon$-*differential privacy*, which works by injecting random noise into the released statistical results computed from the underlying sensitive data, such that the distribution of the noisy results is relatively insensitive to any change of a single record in the original dataset. This ensures that the adversary cannot infer any information about any particular record with high confidence (controlled by parameter $\epsilon$), even if he/she possesses all the remaining tuples of the sensitive data. Meanwhile, the noisy results should be close to the unperturbed ones in order to be useful in practice. Hence, the goal of an $\epsilon$-differential private data publication mechanism is to maximize result accuracy, while satisfying the privacy guarantees.

The best strategy to enforce $\epsilon$-differential privacy depends upon the nature of the statistical analysis that will be performed using the noisy data. This paper focuses on *regression analysis*, which identifies the correlations between different attributes based on the input data. Figure 1 illustrates two most commonly used types of regressions, namely, *linear regression* and *logistic regression*. Specifically, linear regression finds the linear relationship between the input attributes that fits the input data most. In the example shown in Figure 1a, there are two attributes, *age* and *medical expenses*; the data records are shown as dots. The regression result is a straight line with minimum overall distances to the data points, which expresses the value of one attribute as a linear function of the other one. Figure 1b shows an example of logistic regression, where there are two classes of data: diabetes patients (shown as black dots) and those without diabetes (white dots). The goal is to predict the probability of having diabetes, given a patient's other attributes (i.e., age and cholesterol level in our example). The result of this logistic regression can be expressed as a straight line; the probability of a patient getting diabetes is calculated based on which side of the line the patient lies in, and its distance to the line. In particular, if a patient's age and cholesterol level correspond to a point that falls exactly on the straight line, then his/her probability of having diabetes is predicted to be $50\%$. We present the mathematical details of these two types of regression in Section 3.

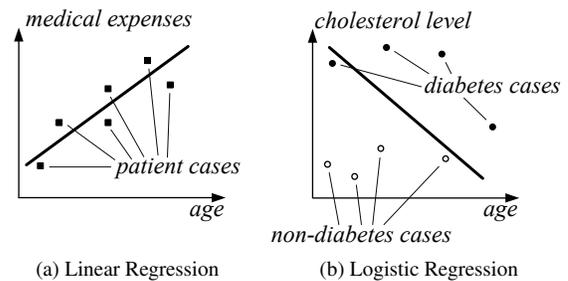

**Figure 1: Two examples of regression problems**

Although regression is a very common type of analysis in practice (especially on medical data), so far there is only a narrow selection of methods for $\epsilon$-differentially private regression. The main challenge lies in the fact that regression involves solving an optimization problem. The relationship between the optimization results and the original data is difficult to analyze; consequently, it is hard to decide on the minimum amount of noise necessary to make the optimization results differentially private. Most existing solutions for $\epsilon$-differentially privacy are designed for releasing simple aggregates (e.g., counts), or structures that can be decomposed into such aggregates, e.g., trees or histograms. One way to adapt these solutions to regression analysis is through *synthetic data generation* (e.g., [7]), which generates synthetic data in a differentially private way based on the original sensitive data. The resulting synthetic dataset can be used for any subsequent analysis. However, due to its generic nature, this methodology often injects an unnecessarily large amount of noise, as shown in our experiments. To our knowledge, the only known solutions that targets regression are [4, 5, 16, 27], which, however, are either limited to non-standard types of regression analysis or unable to produce accurate regression results, as will be shown in Sections 2 and 7.

Motivated by this, we propose the *functional mechanism*, a general framework for enforcing $\epsilon$-differential privacy on analyses that involve solving an optimization problem. The main idea is to enforce $\epsilon$-differential privacy by perturbing the *objective function* of the optimization problem, rather than its results. Publishing the results of the perturbed optimization problem then naturally satisfies $\epsilon$-differential privacy as well. Note that, unlike previous work [4,5] which relies on some special properties of the objective function, our functional mechanism generally applies to all forms of optimization functions. Perturbing objective functions is inherently more challenging than perturbing scalar aggregate values, for two reasons. First, injecting noise into a function is non-trivial; as we show in the paper, simply adding noise to each coefficient of a function often leads to unbearably high noise levels, which in turn leads to nearly useless results. Second, not all noisy functions are valid objective functions; in particular, some noisy functions lead to unbounded results, and some others have multiple local minima. The proposed functional mechanism solves these problems through a set of novel and non-trivial algorithms that perform random perturbations in the functional space. As case studies, we apply the functional mechanism to both linear and logistic regressions. We prove that for both types of regressions, the noise scale required by the proposed methods is constant with respect to the cardinality of the training set. Extensive experiments using real data demonstrate that the functional mechanism achieves highly accurate regression results with comparable prediction power to the unperturbed results, and it significantly outperforms the existing solutions.

The remainder of the paper is organized as follows. Section 2 reviews related studies of differential privacy. Section 3 provides formally defines our problems. Section 4 describes the basic framework for the functional mechanism, and applies it to enforce $\epsilon$-differential privacy on linear regression. Section 5 extends the mechanism to handle more complex objective functions, and solves the problem of differentially private logistic regression. Section 6 presents a post-processing module to ensure that the perturbed objective function has a unique optimal solution. Section 7 contains an extensive set of experimental evaluations. Finally, Section 8 concludes the paper with directions for future work.

## 2. RELATED WORK

Dwork et al. [9] propose $\epsilon$-differential privacy and show that it can be enforced using the *Laplace mechanism*, which supports any queries whose outputs are real numbers (see Section 3 for details). This mechanism is widely adopted in the existing work, but most adoptions are restricted to aggregate queries (e.g., counts) or queries that can be reduced to simple aggregates. In particular, Hay et al. [13], Li et al. [17], Xiao et al. [30], and Cormode et al. [6] present methods for minimizing the worst-case error of a given set of count queries; Barak et al. [2] and Ding et al. [8] consider the publication of data cubes; Xu et al. [31] and Li et al. [18] focus on publishing histograms; McSherry and Mironov [20], Rastogi and Nath [24], and McSherry and Mahajan [19] devise methods for releasing counts on particular types of data, such as time series.

Complement to the Laplace mechanism, McSherry and Talwar [21] propose the *exponential mechanism*, which works for any queries whose output spaces are discrete. This enables differentially private solutions for various interesting problems where the outputs are not real numbers. For instance, the exponential mechanism has been applied for the publication of audition results [21], coresets [10], frequent patterns [3], decision trees [11], support vector machines [25], and synthetic datasets [7, 16]. Nevertheless, neither the Laplace mechanism nor the exponential mechanism can be easily adopted for regression analysis. The reason is that both mechanisms require a careful *sensitivity analysis* of the target problem, i.e, an analysis on how much the problem output would change when an arbitrary tuple in the input data is modified. Unfortunately, such an analysis is rather difficult for regression tasks, due to the complex correlations between regression inputs and outputs.

To the best of our knowledge, the only existing work that targets regression analysis is by Chaudhuri et al. [4, 5], Smith [27], and Lei [16]. Specifically, Chaudhuri et al. [4, 5] show that, when the cost function of a regression task is convex and doubly differentiable, the regression can be performed with a differentially private algorithm based on the objective perturbation. The algorithm, however, is inapplicable for standard logistic regression, as the cost function of logistic regression does not satisfy convexity requirement. Instead, Chaudhuri et al. demonstrate that their algorithm can address a non-standard type of logistic regression with a modified input (see Section 3 for details). Nevertheless, it is unclear whether the modified logistic regression is useful in practice. Smith [27] proposes a general framework for statistical analysis that utilizes both the Laplace mechanism and exponential mechanism. However, the framework requires that the output space of the statistical analysis is bounded, which renders it inapplicable for both linear and logistic regressions. For example, if we preform a linear regression on a three dimensional dataset, the output would be two real numbers (i.e., the slopes of the regression plane on two different dimensions), both of which have an unbounded domain $(-\infty, +\infty)$ (see Section 3 for the details of linear regression).

Lei [16] proposes a regression method that avoids conducting sensitivity analysis directly on the regression outputs. In a nutshell, the method first employs the Laplace mechanism to produce a noisy multi-dimensional histogram of the input data. After that, it produces a synthetic dataset that matches the statistics in the noisy histogram, without looking at the original dataset. Finally, it utilizes the synthetic data to compute the regression results. Observe that, the privacy guarantee of this method is solely decided by the procedure that generates the noisy histogram – the subsequent parts of the algorithm only rely on the histogram (instead of the original data), and hence, they do not reveal any information about the input dataset (except for the information revealed by the noisy histogram). This makes it much easier to enforce $\epsilon$-differential privacy, since the multi-dimensional histogram consists of only counts, which can be processed with the Laplace mechanism in a differentially private manner. Nevertheless, as will be shown in our experiments (referred to as DPME), Lei's method [16] is restricted to datasets with small dimensionality. This is caused by the fact that, when the dimensionality of the input data increases, this method would generate noisy histogram with a coarser granularity, which in turn leads to inaccurate synthetic data and regression results. In summary, none of the existing solutions produce satisfactory results for linear or logistic regressions.

Finally, it is worth mentioning that there exists a relaxed version of $\epsilon$-differential privacy called $(\epsilon, \delta)$-*differential privacy* [23].

Under this privacy notion, a randomized algorithm is considered privacy preserving if it achieves $\epsilon$-differential privacy with a high probability (decided by $\delta$). This relaxed notion is useful in the scenarios where $\epsilon$-differential privacy is too strict to allow any meaningful results to be released (see [12, 15] for examples). As we will show in this paper, however, linear and logistic regressions can be conducted effectively under $\epsilon$-differential privacy, i.e., we do not need to resort to $(\epsilon, \delta)$-differential privacy to achieve meaningful regression results.

## 3. PRELIMINARIES

Let $D$ be a database that contains $n$ tuples $t_1, t_2, \ldots, t_n$ and $d + 1$ attributes $X_1, X_2, \ldots, X_d, Y$. For each tuple $t_i = (x_{i1}, x_{i2}, \ldots, x_{id}, y_i)$, we assume without loss of generality[1] that $\sqrt{\sum_{i=1}^{d} x_{id}^2} \leq 1$.

Our objective is to construct a *regression model* from $D$ that enables us to predict any tuple's value on $Y$ based on its values on $X_1, X_2, \ldots, X_d$, i.e., we aim to obtain a function $\rho$ that (i) takes $(x_{i1}, x_{i2}, \ldots, x_{id})$ as input and (ii) outputs a prediction of $y_i$ that is as accurate as possible.

Depending on the nature of the regression model, the function $\rho$ can be of various types, and it is always parameterized with a vector $\omega$ of real numbers. For example, for linear regression, $\rho$ is a linear function of $x_{i1}, x_{i2}, \ldots, x_{id}$, and the model parameter $\omega$ is a $d$-dimensional vector where the $j$-th ($j \in \{1, \ldots, d\}$) number equals the weight of $x_{ij}$ in the function. To evaluate whether $\omega$ leads to an accurate model, we have a *cost function* $f$ that (i) takes $t_i$ and $\omega$ as input and (ii) outputs a score that measures the difference between the original and predicted values of $y_i$ given $\omega$ as the model parameters. The optimal model parameter $\omega^*$ is defined as:

$$\omega^* = \arg\min_{\omega} \sum_{i=1}^{n} f(t_i, \omega).$$

Without loss of generality, we assume that $\omega$ contains $d$ values $\omega_1, \ldots, \omega_d$. In addition, we consider that $f(t_i, \omega)$ can be written as a function of $\omega_k$ ($k \in \{1, \ldots, d\}$) given $t_i$, as is the case for most regression tasks.

We focus on two commonly-used regression models, namely, linear regression and logistic regression, as defined in the following. For convenience, we abuse notation and use $x_i$ ($i \in \{1, \ldots, d\}$) to denote $(x_{i1}, x_{i2}, \ldots, x_{id})$, and we use $(x_i, y_i)$ to denote $t_i$.

DEFINITION 1 (LINEAR REGRESSION). *Assume without loss of generality that the attribute $Y$ in $D$ has a domain $[-1, 1]$. A linear regression on $D$ returns a prediction function $\rho(x_i) = x_i^T \omega^*$, where $\omega^*$ is a vector of $d$ real numbers that minimizes a cost function $f(t_i, \omega) = \left(y_i - x_i^T \omega\right)^2$, i.e.,*

$$\omega^* = \arg\min_{\omega} \sum_{i=1}^{n} \left(y_i - x_i^T \cdot \omega\right)^2.$$

In other words, linear regression expresses the value of $Y$ as a linear function of the values of $X_1, \ldots, X_d$, such that the sum square error of the predicted $Y$ values is minimized[2].

DEFINITION 2 (LOGISTIC REGRESSION). *Assume that the attribute $Y$ in $D$ has a boolean domain $\{0, 1\}$. A logistic regression on $D$ returns a prediction function, which predicts $y_i = 1$ with probability*

$$\rho(x_i) = \exp(x_i^T \omega^*) / (1 + \exp(x_i^T \omega^*)),$$

*where $\omega^*$ is a vector of $d$ real numbers that minimizes a cost function $f(t_i, \omega) = \log(1 + \exp(x_i^T \omega)) - y_i x_i^T \omega$. That is,*

$$\omega^* = \arg\min_{\omega} \sum_{i=1}^{n} \left(\log(1 + \exp(x_i^T \omega)) - y_i x_i^T \omega\right)$$

For example, assume that $D$ that contains three attributes $X_1$, $X_2$, and $Y$, such that $X_1$ (resp. $X_2$) represents a person's age (resp. body mass index), and $Y$ indicates whether or not the person has diabetes. In that case, a logistic regression on the database would return a function that maps a person's age and body mass index to the probability that he/she would have diabetes, i.e., $\rho(x_i) = Pr[y_i = 1]$. This formulation of logistic regression is used extensively in the medical and social science fields to predict whether certain event will occur given some observed variables. In [4, 5], Chaudhuri et al. consider a non-standard type of logistic regression with modified inputs. In particular, they assume that for each tuple $t_i$, its value on $Y$ is not a boolean value that indicates whether $t_i$ satisfies certain condition; instead, they assume $y_i$ equals the *probability* that a condition is satisfied given $x_i$. For instance, if we are to use Chaudhuri et al.'s method to predict whether a person has diabetes or not based on his/her age and body mass index, then we would need a dataset that gives us accurate likelihood of diabetes for every possible (age, body mass index) combination. This requirement is rather impractical as real datasets are often sparse (due to the curse of dimensionality or the existence of large-domain attributes) and the likelihoods are not measurable. Furthermore, Chaudhuri et al.'s method cannot be applied on datasets where $Y$ is a boolean attribute, since their method relies on convexity property on the cost function. In standard logistic regression, cost function $\log(1 + \exp(x_i^T \omega)) - y_i x_i^T \omega$ (or $\log(1 + \exp(-y_i x_i^T \omega))$ in [4, 5]) does not meet this assumption.

To ensure privacy protection, we require that the regression analysis should be performed with an algorithm that satisfies $\epsilon$-*differential privacy*, which is defined based on the concept of *neighbor databases*, i.e., databases that have the same cardinality but differ in one (and only one) tuple.

DEFINITION 3 ($\epsilon$-DIFFERENTIAL PRIVACY [9]). *A randomized algorithm $\mathcal{A}$ satisfies $\epsilon$-differential privacy, iff for any output $O$ of $\mathcal{A}$ and for any two neighbor databases $D_1$ and $D_2$, we have*

$$Pr\left[\mathcal{A}(D_1) = O\right] \leq e^{\epsilon} \cdot Pr\left[\mathcal{A}(D_2) = O\right].$$

By Definition 3, if an algorithm $\mathcal{A}$ satisfies $\epsilon$-differential privacy for an $\epsilon$ close to 0, then the probability distribution of $\mathcal{A}$'s output is roughly the same for any two input databases that differ in one tuple. This indicates that the output of $\mathcal{A}$ does not reveal significant information about any particular tuple in the input, and hence, privacy is preserved.

As will be shown in Section 4, our solution is built upon the *Laplace mechanism* [9], which is a differentially private framework that can be used to answer any query $Q$ (on $D$) whose output is a vector of real numbers. In particular, the mechanism exploits the *sensitivity* of $Q$, which is defined as

$$S(Q) = \max_{D_1, D_2} \|Q(D_1) - Q(D_2)\|_1, \quad (1)$$

where $D_1$ and $D_2$ are any two neighbor databases, and $\|Q(D_1) - Q(D_2)\|_1$ is the $L_1$ distance between $Q(D_1)$ and $Q(D_2)$. Intuitively, $S(Q)$ captures the maximum changes that could occur in the output of $Q$, when one tuple in the input data is replaced. Given $S(Q)$, the Laplace mechanism ensures $\epsilon$-differential privacy by injecting noise into each value in the output of $Q(D)$, such that the noise $\eta$ follows an *i.i.d. Laplace distribution* with zero mean and scale $S(Q)/\epsilon$ (see [9] for details):

$$pdf(\eta) = \frac{\epsilon}{2S(Q)} \exp\left(-|\eta| \cdot \frac{\epsilon}{S(Q)}\right).$$

---

[1]This assumption can be easily enforced by changing each $x_{ij}$ to $\frac{x_{ij} - \alpha_j}{(\beta_j - \alpha_j) \cdot \sqrt{d}}$, where $\alpha_j$ and $\beta_j$ denotes the minimum and maximum values in the domain of $X_j$.

[2]There is a more general form of linear regression with an objective function $(\omega^*, \alpha^*) = \arg\min_{(\omega, \alpha)} \sum_{i=1}^{n} \left(y_i - x_i^T \cdot \omega - \alpha\right)^2$. We focus only on the type of linear regression in Definition 1 for ease of exposition, but our solution can be easily extended for the more general variant.

In the rest of the paper, we use $Lap(s)$ to denote a random variable drawn from a Laplace distribution with zero mean and scale $s$. For ease of reference, we summarize in Table 1 all notations that will be frequently used.

## 4. FUNCTIONAL MECHANISM

This section presents the *Functional Mechanism (FM)*, a general framework for regression analysis under $\epsilon$-differential privacy. Section 4.1 introduces the details of the framework, while Section 4.2 illustrates how to apply FM on linear regression.

### 4.1 Perturbation of Objective Function

Roughly speaking, FM is an extension of the Laplace mechanism that (i) does not inject noise directly into the regression results, but (ii) ensures privacy by perturbing the optimization goal of regression analysis. To explain, recall that a regression task on a database $D$ returns a model parameter $\omega^*$ that minimizes an optimization function $f_D(\omega) = \sum_{t_i \in D} f(t_i, \omega)$. Direct publication of $\omega^*$ would violate $\epsilon$-differential privacy, since $\omega^*$ reveals information about $f_D(\omega)$ and $D$. One may attempt to address this issue by adding noise to $\omega^*$ using the Laplace mechanism; however, this solution requires an analysis on the sensitivity of $\omega^*$ (see Equation 1), which is rather challenging given the complex correlation between $D$ and $\omega^*$.

Instead of injecting noise directly into $\omega^*$, FM achieves $\epsilon$-differential privacy by (i) perturbing the objective function $f_D(\omega)$ and then (ii) releasing the model parameter $\overline{\omega}$ that minimizes the perturbed objective function $\overline{f}_D(\omega)$ (instead of the original one). A key issue here is: how can we perturb $f_D(\omega)$ in a differentially private manner given that $f_D(\omega)$ can be a complicated function of $\omega$? We address this issue by exploiting the polynomial representation of $f_D(\omega)$, as will be shown in the following.

Recall that $\omega$ is a vector that contains $d$ values $\omega_1, \ldots, \omega_d$. Let $\phi(\omega)$ denote a product of $\omega_1, \ldots, \omega_d$, namely, $\phi(\omega) = \omega_1^{c_1} \cdot \omega_2^{c_2} \ldots \omega_d^{c_d}$ for some $c_1, \ldots, c_d \in \mathbb{N}$. Let $\Phi_j$ ($j \in \mathbb{N}$) denote the set of all products of $\omega_1, \ldots, \omega_d$ with degree $j$, i.e.,

$$\Phi_j = \left\{ \omega_1^{c_1} \omega_2^{c_2} \ldots \omega_d^{c_d} \mid \sum_{l=1}^{d} c_l = j \right\} \quad (2)$$

For example, $\Phi_0 = \{1\}$, $\Phi_1 = \{\omega_1, \ldots, \omega_d\}$, and $\Phi_2 = \{\omega_i \cdot \omega_j \mid i, j \in [1, d]\}$. By the Stone-Weierstrass Theorem [26], any continuous and differentiable $f(t_i, \omega)$ can *always* be written as a (potentially infinite) polynomial of $\omega_1, \ldots, \omega_d$, i.e., for some $J \in [0, \infty]$, we have

$$f(t_i, \omega) = \sum_{j=0}^{J} \sum_{\phi \in \Phi_j} \lambda_{\phi t_i} \phi(\omega), \quad (3)$$

where $\lambda_{\phi t_i} \in \mathbb{R}$ denotes the coefficient of $\phi(\omega)$ in the polynomial. Similarly, $f_D(\omega)$ can also be expressed as a polynomial of $\omega_1, \ldots, \omega_d$.

| Notation | Description |
|---|---|
| $D$ | database of $n$ records |
| $t_i = (x_i, y_i)$ | the $i$-th tuple in $D$ |
| $d$ | the number of values in the vector $x_i$ |
| $\omega$ | the parameter vector of the regression model |
| $f(t_i, \omega)$ | the cost function of the regression model that evaluates whether a model parameter $\omega$ leads to an accurate prediction for a tuple $t_i = (x_i, y_i)$ |
| $f_D(\omega)$ | $f_D(\omega) = \sum_{t_i \in D} f(t_i, \omega)$ |
| $\omega^*$ | $\omega^* = \arg\min_\omega f_D(\omega)$ |
| $\overline{f}_D(\omega)$ | a noisy version of $f_D(\omega)$ |
| $\overline{\omega}$ | $\overline{\omega} = \arg\min_\omega \overline{f}_D(\omega)$ |
| $\tilde{f}_D(\omega)$ | the Taylor expansion of $f_D(\omega)$ |
| $\tilde{\omega}$ | $\tilde{\omega} = \arg\min_\omega \tilde{f}_D(\omega)$ |
| $\hat{f}_D(\omega)$ | a low order approximation of $\tilde{f}_D(\omega)$ |
| $\hat{\omega}$ | $\hat{\omega} = \arg\min_\omega \hat{f}_D(\omega)$ |
| $\phi$ | a product of one or more values in $\omega$, e.g, $(\omega_1)^3 \cdot \omega_2$ |
| $\Phi_j$ | the set of all possible $\phi$ of order $j$ |
| $\lambda_{\phi t_i}$ | the polynomial coefficient of $\phi$ in $f(t_i, \omega)$ |

**Table 1: Table of notations**

Given the above polynomial representation of $f_D(\omega)$, we perturb $f_D(\omega)$ by injecting Laplace noise into its polynomial coefficients, and then derive the model parameter $\overline{\omega}$ that minimizes the perturbed function $\overline{f}_D(\omega)$, as shown in Algorithm 1. The correctness of Algorithm 1 is based on the following lemma and theorem.

LEMMA 1. *Let $D$ and $D'$ be any two neighbor databases. Let $f_D(\omega)$ and $f_{D'}(\omega)$ be the objective functions of regression analysis on $D$ and $D'$, respectively, and denote their polynomial representations as follows:*

$$f_D(\omega) = \sum_{j=1}^{J} \sum_{\phi \in \Phi_j} \sum_{t_i \in D} \lambda_{\phi t_i} \phi(\omega),$$

$$f_{D'}(\omega) = \sum_{j=1}^{J} \sum_{\phi \in \Phi_j} \sum_{t'_i \in D'} \lambda_{\phi t'_i} \phi(\omega).$$

*Then, we have the following inequality:*

$$\sum_{j=1}^{J} \sum_{\phi \in \Phi_j} \left\| \sum_{t_i \in D} \lambda_{\phi t_i} - \sum_{t'_i \in D'} \lambda_{\phi t'_i} \right\|_1 \leq 2 \max_t \sum_{j=1}^{J} \sum_{\phi \in \Phi_j} \|\lambda_{\phi t}\|_1.$$

*where $t_i$ is an arbitrary tuple.*

**Algorithm 1** Functional Mechanism (Database $D$, objective function $f_D(\omega)$, privacy budget $\varepsilon$)
---
1: Set $\Delta = 2\max_t \sum_{j=1}^{J} \sum_{\phi \in \Phi_j} \|\lambda_{\phi t}\|_1$
2: **for** each $0 \leq j \leq J$ **do**
3:    **for** each $\phi \in \Phi_j$ **do**
4:       set $\lambda_\phi = \sum_{t_i \in D} \lambda_{\phi t_i} + Lap\left(\frac{\Delta}{\varepsilon}\right)$
5:    **end for**
6: **end for**
7: Let $\overline{f}_D(\omega) = \sum_{j=1}^{J} \sum_{\phi \in \Phi_j} \lambda_\phi \phi(\omega)$
8: Compute $\overline{\omega} = \arg\min_\omega \overline{f}_D(\omega)$
9: Return $\overline{\omega}$

---

PROOF. Without loss of generality, assume that $D$ and $D'$ differ in the last tuple. Let $t_n$ ($t'_n$ resp.) be the last tuple in $D$ ($D'$ resp.). Then,

$$\sum_{j=1}^{J} \sum_{\phi \in \Phi_j} \left\| \sum_{t_i \in D} \lambda_{\phi t_i} - \sum_{t'_i \in D'} \lambda_{\phi t'_i} \right\|_1 = \sum_{j=1}^{J} \sum_{\phi \in \Phi_j} \|\lambda_{\phi t_n} - \lambda_{\phi t'_n}\|_1$$

$$\leq \sum_{j=1}^{J} \sum_{\phi \in \Phi_j} \|\lambda_{\phi t_n}\|_1 + \sum_{j=1}^{J} \sum_{\phi \in \Phi_j} \|\lambda_{\phi t'_n}\|_1$$

$$\leq 2 \max_t \sum_{j=1}^{J} \sum_{\phi \in \Phi_j} \|\lambda_{\phi t}\|_1$$

$\square$

THEOREM 1. *Algorithm 1 satisfies $\epsilon$-differential privacy.*

PROOF. Let $D$ and $D'$ be two neighbor databases. Without loss of generality, assume that $D$ and $D'$ differ in the last tuple. Let $t_n$ ($t'_n$) be the last tuple in $D$ ($D'$). $\Delta$ is calculated as done on Line 1 of Algorithm 1, and $\overline{f}(\omega) = \sum_{j=1}^{J} \sum_{\phi \in \Phi_j} \lambda_\phi \phi(\omega)$ be the output of Line 7 of the algorithm. We have

$$\frac{Pr\{\overline{f}(\omega) \mid D\}}{Pr\{\overline{f}(\omega) \mid D'\}} = \frac{\prod_{j=1}^{J} \prod_{\phi \in \Phi_j} exp\left(\frac{\epsilon \cdot \|\sum_{t_i \in D} \lambda_{\phi t_i} - \lambda_\phi\|_1}{\Delta}\right)}{\prod_{j=1}^{J} \prod_{\phi \in \Phi_j} exp\left(\frac{\epsilon \cdot \|\sum_{t'_i \in D'} \lambda_{\phi t'_i} - \lambda_\phi\|_1}{\Delta}\right)}$$

$$\leq \prod_{j=1}^{J} \prod_{\phi \in \Phi_j} exp\left(\frac{\epsilon}{\Delta} \cdot \left\| \sum_{t_i \in D} \lambda_{\phi t_i} - \sum_{t'_i \in D'} \lambda_{\phi t'_i} \right\|_1 \right)$$

$$= \prod_{j=1}^{J} \prod_{\phi \in \Phi_j} exp\left(\frac{\epsilon}{\Delta} \cdot \|\lambda_{\phi x_n} - \lambda_{\phi x'_n}\|_1\right)$$

$$= exp\left(\frac{\epsilon}{\Delta} \cdot \sum_{j=1}^{J} \sum_{\phi \in \Phi_j} \|\lambda_{\phi t_n} - \lambda_{\phi t'_n}\|_1\right)$$

$$\leq exp\left(\frac{\epsilon}{\Delta} \cdot 2\max_t \sum_{j=1}^{J} \sum_{\phi \in \Phi_j} \|\lambda_{\phi t}\|_1\right) \quad \text{(by Lemma 1)}$$

$$= exp(\epsilon).$$

In other words, the computation of $\overline{f}(\omega)$ ensures $\epsilon$-differential privacy. The final result of Algorithm 1 is derived from $\overline{f}(\omega)$ without using any additional information from the original database. Therefore, Algorithm 1 is $\epsilon$-differentially private. $\square$

One potential issue with Algorithm 1 is the optimization on the noisy objective function $\overline{f}_D(\omega)$ (see Line 8) can be unbounded when the amount of noise inserted is sufficiently large, leading to meaningless regression results. We address this issue later in Section 6.

In the following, we provide a convergence analysis on Algorithm 1, showing that its output $\overline{\omega}$ is arbitrarily close to the actual minimizer of $f_D(\omega)$, when the database cardinality $n$ is sufficiently large. Our analysis focuses the averaged objective function $\frac{1}{n} f_D(\omega)$ instead of $f_D(\omega)$, since the latter one monotonically increases with $n$. Assume we have a series of databases, $\{D_1, D_2, \ldots, D_n, \ldots\}$, where each $D_j$ contains $j$ tuples all drawn from a fixed but unknown distribution following probability distribution function $p(t)$. We have the following lemma.

LEMMA 2. *If $\lambda_{\phi t}$ is a bounded real number in $(-\infty, +\infty)$ for any $t$ and $\phi \in \cup_{j=1}^{J} \Phi_j$, there exists a polynomial $g(\omega)$ with constant coefficients such that $\lim_{n \to \infty} \frac{1}{n} f_{D_n}(\omega) = g(\omega)$.*

PROOF. Based on our polynomial representation scheme, $\frac{1}{n} f_{D_n}(\omega) = \sum_{j=1}^{J} \sum_{\phi \in \Phi_j} \left(\frac{1}{n} \sum_{i=1}^{n} \lambda_{\phi t_i}\right) \phi(\omega)$, where each $t_i$ is an i.i.d. sample from $p(t)$. When the database cardinality $n$ approaches $+\infty$, we rewrite $\frac{1}{n} \sum_{i=1}^{n} \lambda_{\phi t_i}$ as follows:

$$\lim_{n \to \infty} \frac{1}{n} \sum_{i=1}^{n} \lambda_{\phi t_i} = \int_t \lambda_{\phi t} p(t) dx = E(\lambda_{\phi t}) = c_\phi. \quad (4)$$

By the assumption that $\lambda_{\phi t}$ is bounded, $c_\phi = E(\lambda_{\phi t})$ is a constant that always exists for any $\phi \in \cup_{j=1}^{J} \Phi_j$. Thus, $\lim_{n \to \infty} \frac{1}{n} f_{D_n}(\omega) = \sum_{j=1}^{J} \sum_{\phi \in \Phi_j} c_\phi \phi(\omega)$. This completes the proof, by letting $g(\omega) = \sum_{j=1}^{J} \sum_{\phi \in \Phi_j} c_\phi \phi(\omega)$. $\square$

THEOREM 2. *When database cardinality $n \to +\infty$, the output of Algorithm 1 $\overline{\omega}$ satisfies $g(\overline{\omega}) = \min_\omega g(\omega)$, if $\lambda_{\phi t}$ is bounded for any $t$ and $\phi \in \cup_{j=1}^{J} \Phi_j$.*

PROOF. To prove this theorem, we first show that $\lim_{n \to \infty} \frac{1}{n} \overline{f}_{D_n}(\omega) = g(\omega)$ for any $\omega$.

Given Algorithm 1 with input dataset $D_n$, objective function $f_{D_n}(\omega)$ and privacy budget $\epsilon$, the averaged perturbed objective function $\frac{1}{n} \overline{f}_{D_n}(\omega) = \sum_{j=1}^{J} \sum_{\phi \in \Phi_j} \left(\frac{1}{n} \lambda_\phi\right) \phi(\omega) = \sum_{j=1}^{J} \sum_{\phi \in \Phi_j} \frac{1}{n} \left(\sum_{i=1}^{n} \lambda_{\phi t_i} + Lap(\Delta/\varepsilon)\right) \phi(\omega)$. When $n \to +\infty$, we have

$$\lim_{n \to \infty} \frac{1}{n} \left(\sum_{i=1}^{n} \lambda_{\phi t_i} + Lap\left(\frac{\Delta}{\varepsilon}\right)\right)$$

$$= \lim_{n \to \infty} \frac{1}{n} \sum_{i=1}^{n} \lambda_{\phi t_i} + \lim_{n \to \infty} \frac{1}{n} Lap\left(\frac{\Delta}{\varepsilon}\right)$$

$$= c_\phi + \lim_{n \to \infty} Lap\left(\frac{\Delta}{n\varepsilon}\right),$$

When $\Delta$ and $\epsilon > 0$ are both finite real numbers, it follows that $\lim_{n \to \infty} Lap\left(\frac{\Delta}{n\varepsilon}\right) = 0$. It leads to

$$\lim_{n \to \infty} \frac{1}{n} \overline{f}_{D_n}(\omega) = g(\omega). \quad (5)$$

Since Equation 5 is applicable to any $\omega$, we have $g(\overline{\omega}) = \min_\omega g(\omega)$ by proving $\lim_{n \to \infty} \frac{1}{n} \overline{f}_{D_n}(\overline{\omega}) = \min_\omega \lim_{n \to \infty} \frac{1}{n} \overline{f}_{D_n}(\omega)$, which is obvious given the definition of $\overline{\omega}$ in Algorithm 1. $\square$

Combining the results of Lemma 2 and Theorem 2, we conclude that the output of Algorithm 1 approaches the minimizer of $f_{D_n}(\omega)$, when database cardinality $n \to +\infty$.

## 4.2 Application to Linear Regression

After presenting the general framework, we next apply FM to linear regression as shown in Definition 1. In linear regression, recall that $t_i = (x_i, y_i)$ is the $i$-th tuple in database $D$ with $\sqrt{\sum_{i=1}^{d} x_{id}^2} \leq 1$ and $y_i \in [-1, 1]$; $\omega$ is a $d$-dimensional vector contains model parameters. The expansion of objective function $f_D(\omega)$ for linear regression is

$$f_D(\omega) = \sum_{t_i \in D} \left(y_i - x_i^T \omega\right)^2$$

$$= \sum_{t_i \in D} (y_i)^2 - \sum_{j=1}^{d} \left(2 \sum_{t_i \in D} y_i x_{ij}\right) \omega_j + \sum_{1 \leq j,l \leq d} \left(\sum_{t_i \in D} x_{ij} x_{il}\right) \omega_j \omega_l.$$

Therefore, $f_D(\omega)$ only involves monomials in $\Phi_0$, $\Phi_1$ and $\Phi_2$. Since each $x_i$ locates in the $d$-dimensional unit sphere and $y_i \in [-1, 1]$, given objective function of linear regression, Line 1 of Algorithm 1 could calculate the parameter $\Delta$ as

$$\Delta = 2 \max_{t=(x,y)} \sum_{j=1}^{J} \sum_{\phi \in \Phi_j} \|\lambda_{\phi t}\|_1$$

$$\leq 2 \max_{t=(x,y)} \left(y^2 + 2\sum_{j=1}^{d} y x_{(j)} + \sum_{1 \leq j,l \leq d} x_{(j)} x_{(l)}\right)$$

$$\leq 2(1 + 2d + d^2),$$

where $t$ is an arbitrary tuple and $x_{(j)}$ denotes the $j$-th dimension of vector $x$. Thus, Algorithm 1 adds $Lap(2(d+1)^2/\varepsilon)$ noise to each coefficient and the optimization on $\omega$ is run on the noisy objective function.

For example, assume that we have a two-dimensional database $D$ with three tuples: $(x_1, y_1) = (1, 0.4)$, $(x_2, y_2) = (0.9, 0.3)$, and $(x_3, y_3) = (-0.5, -1)$. The objective function for linear regression is $f_D(\omega) = 2.06\omega^2 - 2.34\omega + 1.25$, with optimal $\omega^* = \frac{117}{206}$. If we apply Algorithm 1 on $D$, then Line 1 of Algorithm 1 would set $\Delta = 2(d+1)^2 = 8$, and then generate the noisy objective function $\overline{f}_D(\omega)$. Figure 2 shows an example of $f_D$ and $\overline{f}_D$. Notice that the global optimum of $\overline{f}_D(\omega)$ is close to the original $\omega^*$ when the coefficients are approximately preserved.

The analysis for linear regression is fairly simple, because the objective function is itself a polynomial on $\omega$. For other regression tasks (e.g., logistic regression), Algorithm 1 cannot be directly applied, as the objective function may not be a polynomial with finite order. In the next section, we will present a solution to tackle this problem.

## 5. POLYNOMIAL APPROXIMATION OF OBJECTIVE FUNCTIONS

For Algorithm 1 to work, it is crucial that the polynomial form of the objective function $f_D(\omega)$ contains only terms with bounded degrees. While this condition holds for certain types of regression analysis (e.g., linear regression, as we have shown in Section 4), there exist regression tasks where the condition cannot be satisfied (e.g., logistic regression). To address this issue, this section presents a method for deriving an *approximate* polynomial form of $f_D(\omega)$ based on *Taylor expansions*. For ease of exposition, we will focus on logistic regression, but our method can be adopted for other types of regression tasks as well.

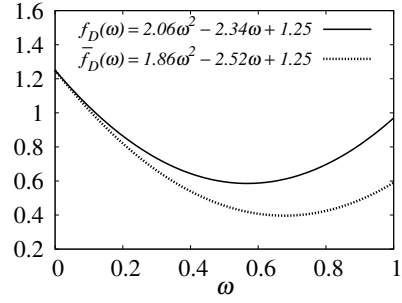

**Figure 2: Example of objective function for linear regression and its noisy version obtained by FM**

## 5.1 Expansion

Consider the cost function $f(t_i, \omega)$ of regression analysis. Assume that there exist $2m$ functions $f_1, \ldots, f_m$ and $g_1, \ldots, g_m$, such that (i) $f(t_i, \omega) = \sum_{l=1}^{m} f_l(g_l(t_i, \omega))$, and (ii) each $g_l$ is a polynomial function of $\omega_1, \ldots, \omega_m$. (As will be shown shortly, such a decomposition of $f(t_i, \omega)$ is useful for handling logistic regression.) Given the above decomposition of $f(t_i, \omega)$, we can apply the Taylor expansion on each $f_l(\cdot)$ to obtain the following equation:

$$\tilde{f}(t_i, \omega) = \sum_{l=1}^{m} \sum_{k=0}^{\infty} \frac{f_l^{(k)}(z_l)}{k!} \left(g_l(t_i, \omega) - z_l\right)^k, \quad (6)$$

where each $z_l$ is a real number. Accordingly, the objective function $f_D(\omega)$ can be written as:

$$\tilde{f}_D(\omega) = \sum_{i=1}^{n} \sum_{l=1}^{m} \sum_{k=0}^{\infty} \frac{f_l^{(k)}(z_l)}{k!} \left(g_l(t_i, \omega) - z_l\right)^k \quad (7)$$

To explain how Equations 6 and 7 are related to logistic regression, recall that the cost function of logistic regression is $f(t_i, \omega) = \log(1 + \exp(x_i^T \omega)) - y_i x_i^T \omega$. Let $f_1$, $f_2$, $g_1$, and $g_2$ be four functions defined as follows:

$$g_1(t_i, \omega) = x_i^T \omega, \qquad g_2(t_i, \omega) = y_i x_i^T \omega,$$
$$f_1(z) = \log(1 + \exp(z)), \qquad f_2(z) = z.$$

Then, we have $f(t_i, \omega) = f_1(g_1(t_i, \omega)) + f_2(g_2(t_i, \omega))$. By Equations 6 and 7,

$$\tilde{f}_D(\omega) = \sum_{i=1}^{n} \sum_{l=1}^{2} \sum_{k=0}^{\infty} \frac{f_l^{(k)}(z_l)}{k!} \left(g_l(t_i, \omega) - z_l\right)^k \quad (8)$$

Since $f_2(z) = z$, we have $f_2^{(k)} = 0$ for any $k > 1$. Given this fact and by setting $z_l = 0$, Equation 8 can be simplified as

$$\tilde{f}_D(\omega) = \sum_{i=1}^{n} \sum_{k=0}^{\infty} \frac{f_1^{(k)}(0)}{k!} \left(x_i^T \omega\right)^k - \left(\sum_{i=1}^{n} y_i x_i^T\right) \omega \quad (9)$$

There are two complications in Equation 9 that prevent us from applying it for private logistic regression. First, the equation involves a infinite summation. Second, the term $f_1^{(k)}(0)$ involved in the equation does not have closed form solution. To address these two issues, we will present an approximate approach that reduces the degree of the summation, and the approach only requires the value of $f_1^{(k)}(0)$ for $k = 0, 1, 2$, i.e., $f_1^{(0)}(0) = \log 2$, $f_1^{(1)}(0) = \frac{1}{2}$, and $f_1^{(2)}(0) = \frac{1}{4}$.

## 5.2 Approximation

Our approximation approach works by truncating the Taylor series in Equation 9 to remove all polynomial terms with order larger than 2. This leads to a new objective function with only low order polynomials as follows:

$$\hat{f}_D(\omega) = \sum_{l=1}^{m} \sum_{i=1}^{n} \hat{f}_l(g_l(t_i, \omega))$$
$$= \sum_{l=1}^{m} \sum_{i=1}^{n} \sum_{k=0}^{2} \frac{f_l^{(k)}(z_l)}{k!} (g_l(t_i, \omega) - z_l)^k \quad (10)$$

A natural question is: how much error would the above approximation approach incur? The following lemmata provide the answer.

LEMMA 3. *Let* $\tilde{\omega} = \arg\min_\omega \tilde{f}_D(\omega)$ *and* $\hat{\omega} = \arg\min_\omega \hat{f}_D(\omega)$. *Let* $L = \max_\omega \left( \tilde{f}_D(\omega) - \hat{f}_D(\omega) \right)$ *and* $S = \min_\omega \left( \tilde{f}_D(\omega) - \hat{f}_D(\omega) \right)$. *We have the following inequality:*

$$\tilde{f}_D(\hat{\omega}) - \tilde{f}_D(\tilde{\omega}) \leq L - S \quad (11)$$

PROOF. Observe that $L \geq \tilde{f}_D(\hat{\omega}) - \hat{f}_D(\hat{\omega})$ and $S \leq \tilde{f}_D(\omega^*) - \hat{f}_D(\omega^*)$. Therefore,

$$\tilde{f}_D(\hat{\omega}) - \hat{f}_D(\hat{\omega}) - \tilde{f}_D(\omega^*) + \hat{f}_D(\omega^*) \leq L - S.$$

In addition, $\hat{f}_D(\hat{\omega}) - \hat{f}_D(\omega^*) \leq 0$. Hence, Equation 11 holds. □

Lemma 3 shows that the error incurred by truncating the Taylor series approximate function depends on the maximum and minimum values of $\tilde{f}_D(\omega) - \hat{f}_D(\omega)$. To quantify the magnitude of the error, we first rewrite $\tilde{f}_D(\omega) - \hat{f}_D(\omega)$ in a form similar to Equation 8:

$$\tilde{f}_D(\omega) - \hat{f}_D(\omega) = \sum_{l=1}^{m} \sum_{i=1}^{n} \sum_{k=3}^{\infty} \frac{f_l^{(k)}(z_l)}{k!} (g_l(t_i, \omega) - z_l)^k$$

To derive the minimum and maximum values of the function above, we look into the remainder of Taylor expansion. The following lemma provides exact lower and upper bounds on $\tilde{f}_D(\omega) - \hat{f}_D(\omega)$, which is a well known result [1].

LEMMA 4. *For any* $z \in [z_l - 1, z_l + 1]$, $\frac{1}{n} \left( \tilde{f}_D(\omega) - \hat{f}_D(\omega) \right)$ *must be in the interval*

$$\left[ \sum_l \frac{\min f_l^{(3)}(z)(z - z_l)^3}{6}, \sum_l \frac{\max f_l^{(3)}(z)(z - z_l)^3}{6} \right]$$

By combining Lemmata 3 and 4, we can easily calculate the error incurred by our approximation approach. In particular, the error only depends on the structure of the function, and is independent of the characteristics of the dataset. Furthermore, the average error of the approximation is always bounded, since

$$\frac{1}{n}\tilde{f}_D(\omega) - \frac{1}{n}\hat{f}_D(\omega)$$
$$\leq \sum_l \frac{\max_z f^{(3)}(z)(z - z_l)^3 - \min_z f^{(3)}(z)(z - z_l)^3}{6}.$$

The above analysis applies to the case of logistic regression as follows. First, for the function $f_1(z) = \log(1 + \exp(z))$, we have $f_1^{(3)}(z) = \frac{\exp(z) - (\exp(z))^2}{(1+\exp(z))^3}$. It can be verified that $\min_z f_1^{(3)}(z) = \frac{e - e^2}{(1+e)^3}$, $\max_z f_1^{(3)}(z) = \frac{e^2 - e}{(1+e)^3}$. Thus, the average error of the approximation is at most

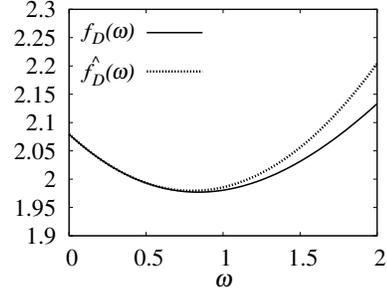

Figure 3: Example of objective function for logistic regression and its polynomial approximation

$$\tilde{f}(\hat{\omega}) - \tilde{f}(\tilde{\omega}) \leq \frac{(e^2 - e)}{6(1+e)^3} - \frac{(e - e^2)}{6(1+e)^3}$$
$$= \frac{(e^2 - e)}{6(1+e)^3}$$
$$\approx 0.015$$

In other words, the error of the approximation on logistic regression is a small constant. However, because of this error, there does not exist a convergence result similar to the one stated in Theorem 2. That is, there is a gap between the results from our approximation approach and those from a standard regression algorithm. To illustrate this, let us consider a two-dimensional database $D$ with three tuples, $(x_1, y_1) = (-0.5, 1)$, $(x_2, y_2) = (0, 0)$, and $(x_3, y_3) = (1, 1)$. Figure 3 illustrates the objective function of logistic regression $f_D$ as well as its approximation $\hat{f}_D$. As will be shown in our experiments, however, our approximation approach still leads to accuracy regression results.

## 5.3 Application to Logistic Regression

Algorithm 2 presents an extension of Algorithm 1 that incorporates our polynomial approximation approach. In particular, given a dataset $D$, Algorithm 2 first constructs a new objective function $\hat{f}_D(\omega)$ that approximates the original one, and then feeds the new objective function as input to Algorithm 1. The model parameter returned from Algorithm 1 is then output as the final results of Algorithm 2. It can be verified that Algorithm 2 guarantees $\epsilon$-differential privacy – this follows from the fact that (i) Algorithm 1 guarantees $\epsilon$-differential privacy for any given objective function (regardless of whether it is an approximation), and (ii) the output of Algorithm 2 is directly obtained from Algorithm 1.

To apply Algorithm 2 for logistic regression, we set

$$\hat{f}_D(\omega) = \sum_{i=1}^{n} \sum_{k=0}^{2} \frac{f_1^{(k)}(0)}{k!} \left( x_i^T \omega \right)^k - \left( \sum_{i=1}^{n} y_i x_i^T \right) \omega.$$

(This is by Equation 8 and the fact that $\hat{f}_D(\omega)$ retains only the low order terms in the $f_D(\omega)$.) After that, when $\hat{f}_D(\omega)$ is fed as part of input to Algorithm 1, Line 1 of Algorithm 1 would calculate the

parameter $\Delta$ as:

$$\begin{aligned}
\Delta &= 2 \max_{t=(x,y)} \left( \frac{f_1^{(1)}(0)}{1!} \sum_{j=1}^{d} x_{(j)} + \frac{f_1^{(2)}(0)}{2!} \sum_{j,l} x_{(j)} x_{(l)} \right. \\
&\quad \left. + y \sum_{j=1}^{d} x_{(j)} \right) \\
&\leq 2(\frac{d}{2} + \frac{d^2}{8} + d) \\
&= \frac{d^2}{4} + 3d,
\end{aligned}$$

where $t$ is an arbitrary tuple and $x_{(j)}$ denotes the $j$-th dimension of the vector $x$.

Recall that Algorithm 1 injects Laplace noise with scale $\Delta/\epsilon$ to the coefficients of the objective function (see Line 4 of Algorithm 1). Therefore, $\Delta = d^2/4 + 3d$ indicates that the amount of noise injected by our algorithm is only related to $d$ and is independent of the dataset cardinality.

## 6. AVOIDING UNBOUNDED NOISY OBJECTIVE FUNCTIONS

As shown in the previous sections, FM achieves $\epsilon$-differential privacy by injecting Laplace noise into the coefficients of the objective functions of optimization problems. The injection of noise, however, may render the objective function *unbounded*, i.e., there may not exist any optimal solution for the noisy objective function. For instance, if we fit a linear regression model on a two dimensional dataset, the objective function would be a quadratic function $f_D(\omega) = a\omega^2 + b\omega + c$ with a minimum point (see Figure 2 for an example). If we add noise into the coefficients of $f_D(\omega)$, however, the resulting objective function may be no longer have a minimum, i.e., when coefficient $a$ becomes non-positive after noise injection. In that case, there does not exist a solution to the optimization problem.

One simple approach to address the above issue is to re-run FM whenever the noisy objective function is unbounded, until we obtain a solution to the optimization problem. This approach, as shown in the following lemma, ensures $\epsilon$-differential privacy but incurs two times the privacy cost of FM.

LEMMA 5. *Let $\mathcal{A}^*$ be an algorithm that repeats Algorithm 1 with privacy budget $\epsilon$ on a dataset, until the output of Algorithm 1 corresponds to a bounded objective function. Then, $\mathcal{A}^*$ satisfies $(2\epsilon)$-differential privacy.*

PROOF. Let $D$ and $D'$ be any two neighbor datasets, $\mathcal{A}$ be Algorithm 1, and $O$ be any output of $\mathcal{A}$. Since $\mathcal{A}$ ensures $\epsilon$-differential privacy (see Theorem 1), we have

$$\begin{aligned}
e^{-\epsilon} \cdot Pr\left[\mathcal{A}(D') = O\right] &\leq Pr\left[\mathcal{A}(D) = O\right] \\
&\leq e^{\epsilon} \cdot Pr\left[\mathcal{A}(D') = O\right] \quad (12)
\end{aligned}$$

Let $\mathcal{O}^+$ be the set of outputs by Algorithm 1 that correspond to bounded objective functions. For any $O^+ \in \mathcal{O}^+$, we have

$$\begin{aligned}
Pr\left[\mathcal{A}^*(D) = O^+\right] &= \frac{Pr\left[\mathcal{A}^*(D) = O^+\right]}{\sum_{O' \in \mathcal{O}^+} Pr\left[\mathcal{A}^*(D) = O'\right]} \\
&\leq \frac{e^{\epsilon} \cdot Pr\left[\mathcal{A}^*(D') = O^+\right]}{e^{-\epsilon} \cdot \sum_{O' \in \mathcal{O}^+} Pr\left[\mathcal{A}^*(D') = O'\right]} \quad \text{(By Eqn. 12)} \\
&\leq e^{2\epsilon} \cdot Pr\left[\mathcal{A}^*(D') = O^+\right].
\end{aligned}$$

□

---

**Algorithm 2** Functional Mechanism (Database $D$, objective function $f_D(\omega)$, privacy budget $\varepsilon$)
1: Decompose the function $f(t_i, \omega) = \sum_l f_l(g_l(t_i, \omega))$.
2: Build a new objective function $\hat{f}_D(\omega)$, such that
   $\hat{f}_D(\omega) = \sum_{l=1}^{m} \sum_{i=1}^{n} \sum_{k=0}^{2} \frac{f_l^{(k)}(z_l)}{k!} (g_l(t_i, \omega) - z_l)^k$
3: Run Algorithm 1 with input $\left(D, \hat{f}_D(\omega), \varepsilon\right)$.
4: Return $\overline{\omega}$ from Algorithm 1.

---

Although repeating FM provides a quick fix to obtain bounded objective functions, it leads to sub-optimal results as it entails a considerably higher privacy cost than FM does. To address this issue, we propose two methods to avoid unbounded objective functions in linear and logistic regressions, as will be detailed in Sections 6.1 and 6.2.

### 6.1 Regularization

As shown in Sections 4 and 5, given a linear or logistic regression task, FM would transform the objective function into a quadratic polynomial $\hat{f}_D(\omega)$, after which it injects noise into the coefficients of $\hat{f}_D(\omega)$ to ensure privacy. Let $\hat{f}_D(\omega) = \omega^T M \omega + \alpha \omega + \beta$ be the matrix representation of the quadratic polynomial, and $\bar{f}_D(\omega) = \omega^T M^* \omega + \alpha^* \omega + \beta^*$ be the noisy version of $\hat{f}_D(\omega)$ after injection of Laplace noise. Then, $M$ must be symmetric and positive definite [28]. To ensure that $\hat{f}_D(\omega)$ is bounded after noise injection, it suffices to make $M^*$ also symmetric and positive definite [28].

The symmetry of $M^*$ can be easily achieved by (i) adding noise to the upper triangular part of the matrix and (ii) copying each entry to its counterpart in the lower triangular part. In contrast, it is rather challenging to ensure that $M^*$ is positive definite. To our knowledge, there is no existing method for transforming a positive definite matrix into another positive definite matrix in a differentially private manner. To circumvent this, we adopt a heuristic approach called *regularization* from the literature of regression analysis [14, 29]. In particular, we add a positive constant $\lambda$ to each entry in the main diagonal of $M^*$, such that the noisy objective function becomes

$$\bar{f}_D(\omega) = \omega^T \left(M^* + \lambda I\right) \omega + \alpha^* \omega + \beta^*, \quad (13)$$

where $I$ is a $d \times d$ identity matrix, and $\alpha^*$ and $\beta^*$ are the noisy versions of $\alpha$ and $\beta$, respectively.

Although regularization is mostly used in regression analysis to avoid overfitting [14, 29], it also helps achieving a bounded $\bar{f}_D(\omega)$. To illustrate this, consider that we perform linear regression on a two dimensional database. We have $d = 1$. In addition, each of $\omega$, $M^* + \lambda \cdot I$, $\alpha^*$, and $\beta^*$ contains only one value (see Figure 2 for an example). Accordingly, the noisy objective function $\bar{f}_D(\omega)$ would be a quadratic function with one variable $\omega$. Such a function has a minimum, if and only if $M^* + \lambda I$ is positive. Intuitively, we can ensure this as long as $\lambda$ is large enough to mitigate the noise injected in $M^*$.

In general, for any $d \geq 1$, a reasonably large $\lambda$ makes it more likely that all eigenvalues of $M^* + \lambda I$ are positive, in which case $M^* + \lambda I$ would be positive definite. Meanwhile, as long as $\lambda$ does not overwhelm the signal in $M^*$, it would not significantly degrade the quality of the solution to the regression problem. In our experiments, we observe that a good choice of $\lambda$ equals 4 times standard deviation of the Laplace noise added into $M^*$. Note that setting $\lambda$ to this value does not degrade the privacy guarantee of FM, since the standard deviation of the Laplace noise does not reveal any information about the original dataset.

Although regularization increases the chance of obtaining a bounded objective function, there is still a certain probability that the noise objective function does not have a minimum even after regularization. This motivates our second approach, *spectral trimming*, as will be explained in Section 6.2.

## 6.2 Spectral Trimming

Let $\bar{f}_D(\omega) = \omega^T (M^* + \lambda I) \omega + \alpha^* \omega + \beta^*$ be the noisy objective function with regularization. As we have discussed in Section 6.1, $M^* + \lambda I$ is symmetric (due to the symmetry of $M^*$). In addition, $\bar{f}_D(\omega)$ is unbounded if and only if $M^* + \lambda I$ is not positive definite, which holds if and only if at least one eigenvalue of $M^* + \lambda I$ is not positive [28]. In other words, to transform an unbounded $\bar{f}_D$ into a bounded one, it suffices to get rid of the non-positive eigenvalues of $M^* + \lambda I$.

Let $Q^T \Lambda Q$ be the eigen-decomposition of $M^* + \lambda I$, i.e., $Q$ is a $d \times d$ matrix where each row is an eigenvector of $M^* + \lambda I$, and $\Lambda$ is a diagonal matrix where the $i$-th diagonal element is the eigenvalue of $M^* + \lambda I$ corresponding to the eigenvector in the $i$-th row of $Q$. We have $Q^T Q = I$. Accordingly,

$$\bar{f}_D(\omega) = \omega^T \left(Q^T \Lambda Q\right) \omega + \alpha^* \left(Q^T Q\right) \omega + \beta^*$$

Suppose that the $i$-th diagonal element $e_i$ of $\Lambda$ is not positive. Then, we would remove $e_i$ from $\Lambda$, which results in a $(d-1) \times (d-1)$ diagonal matrix. In addition, we would also delete the $i$ row in $Q$, so that $Q^T \Lambda Q$ would still be well-defined. In general, if $\Lambda$ contains $k$ non-positive diagonal elements, then removal of all those elements would transform $\Lambda$ into a $(d-k) \times (d-k)$ matrix, which we denote as $\Lambda'$. Accordingly, $Q$ becomes a $(d-k) \times d$ matrix, which we denote as $Q'$. The noisy objective function then becomes

$$\bar{f}_D(\omega) = \omega^T \left(Q'^T \Lambda' Q'\right) \omega + \alpha^* \left(Q'^T Q'\right) \omega + \beta^*. \quad (14)$$

We rewrite $\bar{f}_D(\omega)$ as a function of $Q'\omega$:

$$\bar{g}_D\left(Q'\omega\right) = (Q'\omega)^T \Lambda' (Q'\omega) + \alpha^* Q'^T (Q'\omega) + \beta^*,$$

which is a bounded function of $Q'\omega$ since all eigenvalues of $\Lambda'$ are positive. We compute the vector $V$ that minimizes $\bar{g}_D(V)$, and then derive $\omega$ by solving $Q'\omega = V$ (note that the solution to this equation is not unique).

In summary, we delete non-positive elements in $\Lambda$ to obtain a bounded objective function, based on which we derive the model parameters. Intuitively, the non-positive elements in $\Lambda$ are mostly due to noise, and hence, removing them from $\Lambda$ would not incur significant loss of useful information. Therefore, the objective function in Equation 14 may still lead to accurate model parameters. The removal of non-positive elements from $\Lambda$ does not violate $\epsilon$-differential privacy, as the removing procedure depends only on $M^*$ (which is differentially private) instead of the input database.

## 7. EXPERIMENTS

This section experimentally evaluates the performance of FM against four approaches, namely, *DPME* [16], *Filter-Priority (FP)* [7], *NoPrivacy*, and *Truncated*. As explained in Section 2, DPME is the state-of-the-art method for regression analysis under $\epsilon$-differential privacy, while FP is an $\epsilon$-differentially private technique for generating synthetic data that can also be used for regression tasks. *NoPrivacy* and *Truncated* are two algorithms that performs regression analysis do not enforce $\epsilon$-differential privacy: *NoPrivacy* directly outputs the model parameters that minimize the objective function, and *Truncated* returns the parameters obtained from an approximate objective function with truncated polynomial terms (see Section 5). We include *Truncated* in the experiments, so as to investigate the error incurred by the low-order approximation approach proposed in Section 5. For DPME and FP, we use the implementations provided by their respective authors, and we set all internal parameters (e.g., the granularity of noisy histograms used by DPME) to their recommended values. All experiments are conducted using Matlab (version 7.12) on a computer with a 2.4GHz CPU and 32GB RAM.

| Parameter | Range and Default Value |
|---|---|
| Data Subset Sampling Rate | 0.1, 0.2, 0.3, 0.4, 0.5, 0.6, 0.7, 0.8, 0.9, **1** |
| Dataset Dimensionality | 5, 8, **11**, 14 |
| Privacy Budget $\epsilon$ | 3.2, 1.6, **0.8**, 0.4, 0.2, 0.1 |

Table 2: Experimental parameters and values

We use two datasets from the *Integrated Public Use Microdata Series* [22], *US* and *Brazil*, which contain $370,000$ and $190,000$ census records collected in the US and Brazil, respectively. There are $13$ attributes in each datasets, namely, *Age*, *Gender*, *Martial Status*, *Education*, *Disability*, *Nativity*, *Working Hours per Week*, *Number of Years Residing in the Current Location*, *Ownership of Dwelling*, *Family Size*, *Number of Children*, *Number of Automobiles*, and *Annual Income*. Among these attributes, *Marital status* is the only categorical attribute whose domain contains more than $2$ values, i.e., *Single*, *Married*, and *Divorced/Widowed*. Following common practice in regression analysis, we transform *Marital Status* into two binary attributes, *Is Single* and *Is Married* (an individual divorced or widowed would have *false* on both of these attributes). With this transformation, both of our datasets become $14$ dimensional.

We conduct regression analysis on each dataset to predict the value of *Annual Income* using the remaining attributes. For logistic regression, we convert *Annual Income* into a binary attribute: values higher than a predefined threshold are mapped to $1$, and $0$ otherwise. Accordingly, when we use a logistic model to classify a tuple $t$, we predict the *Annual Income* of $t$ to be $1$ if $\frac{\exp(x_i^T \omega)}{1+\exp(x_i^T \omega)} > 0.5$ (see Definition 2), where $\omega$ is the model parameter, and $x$ is a vector that contains the values of $t$ on all attributes expect *Annual Income*. We measure the accuracy of a logistic model by its *misclassification rate*, i.e., the fraction of tuples that are incorrectly classified. The accuracy of a linear model, on the other hand, is measured by the mean square error of the predicted values, i.e., $\frac{1}{n} \sum_{i=1}^n \left(y_i - x_i^T \omega\right)^2$, where $n$ is the number of tuples in the dataset, $y_i$ is the *Annual Income* value of the $i$-th tuple, $x$ is a vector that contains the other attribute values of the tuple, and $\omega$ is the model parameter.

In each experiment, we perform 5-fold cross-validation 50 times for each algorithm, and we report the average results. We vary three different parameters, i.e., the dataset size, the dataset dimensionality, and privacy budget $\epsilon$. In particular, we generate random subsets of the tuples in the US and Brail datasets, with the sampling rate varying from $0.1$ to $1$. To vary the dataset dimensionality, we select three subsets of the attributes in each dataset for classification. The first subset contains $5$ attributes: *Age*, *Gender*, *Education*, *Family Size*, and *Annual Income*. The second subset consists of $8$ attributes: the aforementioned five attributes, as well as *Nativity*, *Ownership of Dwelling*, and *Number of Automobiles*. The third subset contains all attributes in the second subset, as well as *Is Single*, *Is Married*, and *Number of Children*. Table 2 summarizes the parameter values, with the default values in bold.

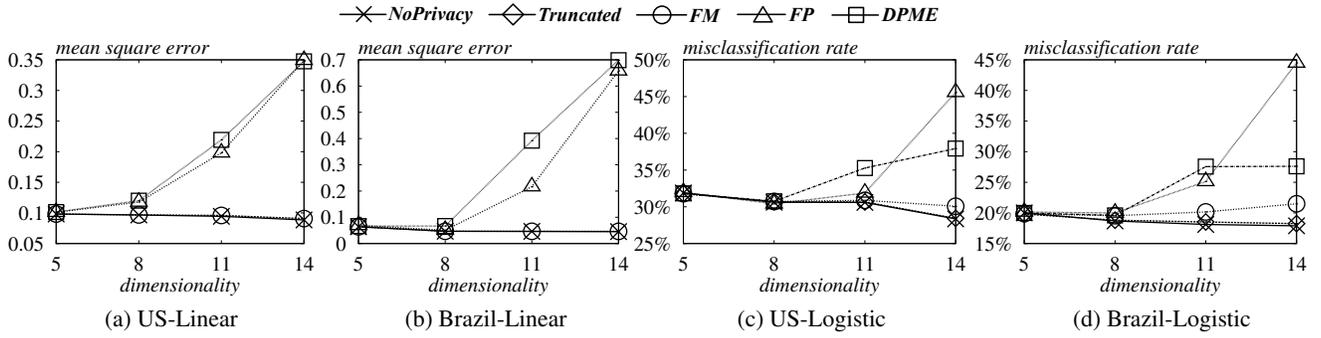

Figure 4: Regression accuracy v.s. dataset dimensionality

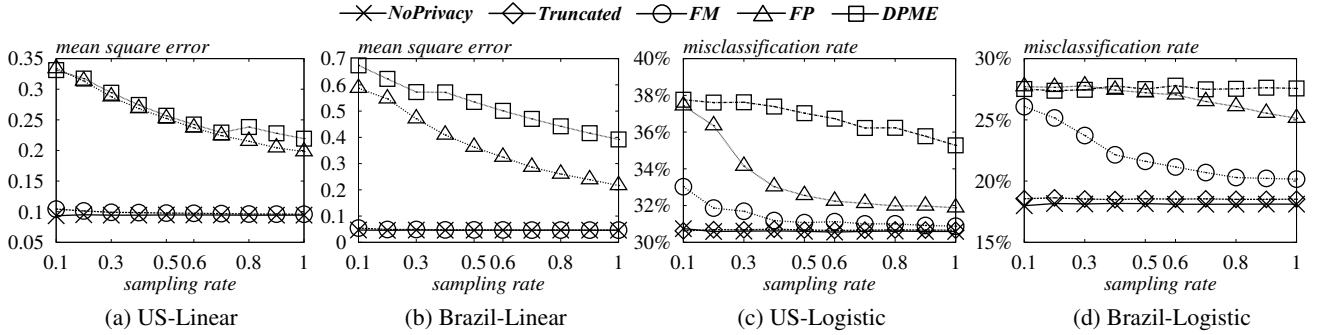

Figure 5: Regression accuracy v.s. dataset cardinality

## 7.1 Accuracy vs. Dataset Dimensionality

Figures 4a and 4b illustrate the linear regression error of each algorithm as a function of the dataset dimensionality. We omit Truncated in the figures, as our approximation approach in Section 5 is required only for logistic regression but not linear regression. Observe that FM consistently outperforms FP and DPME, and its regression accuracy is almost identical to that of of NoPrivacy. In contrast, FP and DPME incur significant errors, especially when the dataset dimensionality is large.

Figures 4c and 4d show the error of each algorithm for logistic regression. The error of Truncated is comparable to that of No-Privacy, which demonstrates the effectiveness of our low-order approximation approach that truncates the polynomial representation of the objective function. The error of FM is slightly higher than that of Truncated , but it is still much smaller than the errors of FP and DPME.

## 7.2 Accuracy vs. Dataset Cardinality

Figure 5 show the regression error of each algorithm as a function of the dataset cardinality. For both regression tasks and for both datasets, FM outperforms FP and DPME by considerable margins. In addition, for linear regression, the difference in accuracy between FM and NoPrivacy is negligible; meanwhile, their accuracy remains stable with varying number of records in the database, except when the sampling rate equals 0.1 (the smallest value used in all experiments). In contrast, the performance of FP and DPME improves with the dataset cardinality, which is consistent with the theoretical result in [7] and [16]. Nevertheless, even when we use all tuples in the dataset, the accuracy of FP and DPME is still much worse than that of FM and NoPrivacy.

For logistic regression, there is a gap between the accuracy of FM and that of NoPrivacy and Truncated, but the gap shrinks rapidly with the increase of dataset cardinality. The errors of FP and DPME also decrease when the dataset cardinality increases, but they remain considerably higher than the error of FM in all cases.

## 7.3 Accuracy vs. Privacy Budget

Figure 6 plots the regression error of each algorithm as a function of the privacy budget $\epsilon$. The errors of NoPrivacy and Truncated remain unchanged for all $\epsilon$, as none of them enforces $\epsilon$-differential privacy. All of the other three methods incur higher errors when $\epsilon$ decreases, as a smaller $\epsilon$ requires a larger amount of noise to be injected. FM outperforms FP and DPME in all cases, and it is relatively robust against the change of $\epsilon$. In contrast, FP and DPME produce much less accurate regression results, especially when $\epsilon$ is small.

## 7.4 Computation Time

Finally, Figures 7-9 report the average running time of each algorithm. Due to the space constraint, we only report the results for logistic regression; the results for linear regression are qualitatively similar. Overall speaking, the running time of FM is at least one order of magnitude lower than that of NoPrivacy, which in turn is about two times faster than FP and DPME. The efficiency of FM is mainly due to its low-order approximation module, which truncates the polynomial representation of the objective function and retains only the first and second order terms. As a consequence, FM computes the optimization results by solving a multi-variate *quadratic optimization* problem, for which Matlab has an efficient solution. In contrast, all other methods require solving the original optimization problem of logistic regression, which has a complicated objective function that renders the solving process time consuming. In addition, FP and DPME require additional time to generate synthetic data, leading to even higher computation cost.

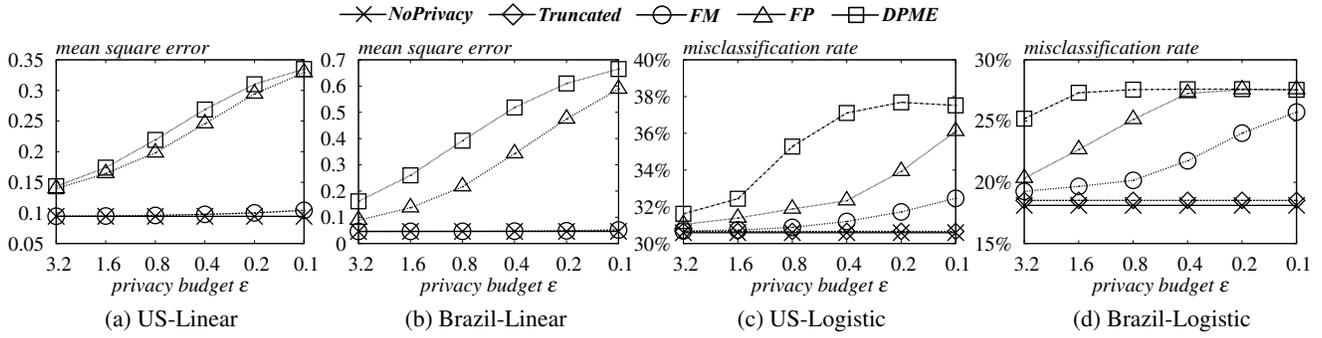

(a) US-Linear    (b) Brazil-Linear    (c) US-Logistic    (d) Brazil-Logistic

Figure 6: Regression accuracy v.s. privacy budget

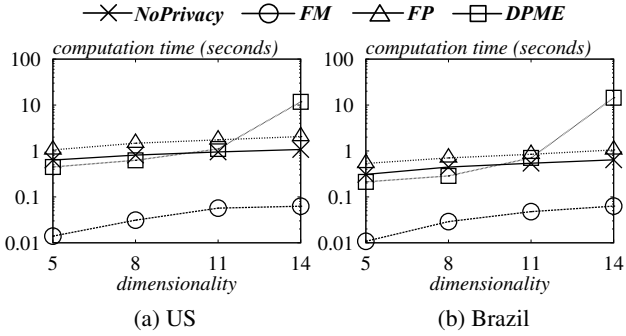

(a) US    (b) Brazil

Figure 7: Computation time v.s. dataset dimensionality on logistic regression

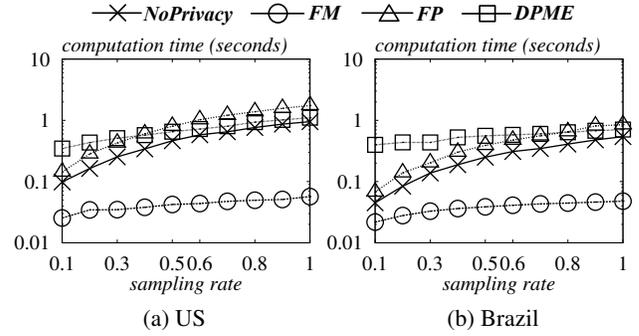

(a) US    (b) Brazil

Figure 8: Computation time v.s. dataset cardinality on logistic regression

As shown in Figure 7 and Figure 8, the computation time of all algorithms increases with the dimensionality and cardinality of the databset. This is expected, as a larger number of tuples (attributes) leads to a higher complexity of the optimization problem. The execution time of FP and DPME increases at a faster rate than that of FM and NoPrivacy, since the former two require generating synthetic data, which entails higher computation cost when the number of tuples (attributes) in the dataset increases. On the other hand, as shown in Figure 9, the privacy budget $\epsilon$ has negligible effects on the running time of the algorithms, since it affects neither the size or dimensionality of the dataset nor the complexity of the optimization problem being solved.

In summary, FM is superior to FP and DPME in terms of both accuracy and efficiency in all experiments. The advantage of FM in terms of regression accuracy is more pronounced when the data dimensionality increases. Furthermore, the accuracy of FM is even comparable to NoPrivacy in the scenarios where the cardinality of the dataset or the privacy budget is reasonably large. These results demonstrate that FM is a preferable method for differentially private regression analysis.

## 8. CONCLUSION AND FUTURE WORK

This paper presents a general approach for differentially private regression. Different from existing techniques, our approach conducts both sensitivity analysis and noise insertion on the objective functions, which leads to more accurate regression results when the objective functions can be represented as finite polynomials. To tackle more complex objective functions with infinite polynomial representation (e.g., logistic regression), we propose to truncate the Taylor expansion of the objective function, and we analyze the error incurred in the optimization results. Our empirical studies on real datasets validate our theoretical results and demonstrate the effectiveness and efficiency of our proposal.

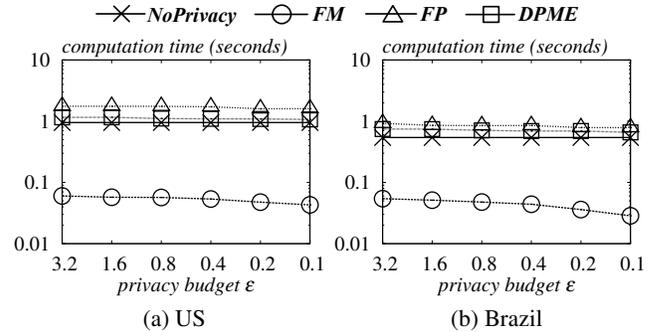

(a) US    (b) Brazil

Figure 9: Computation time v.s. privacy budget on logistic regression

For future work, we plan to extend our research on the following directions. First, our current mechanism only works with objective functions in the form of $\sum_{i=1}^{n} f(t_i, \omega)$. However, there exist regression tasks with more complicated objective functions (e.g., Cox regression). It is interesting and challenging to investigate how those regression tasks can be addressed. Second, besides Taylor expansion, there may exist other analytical tool that can be used to approximate the objective functions. We plan to study whether alternative analytical tool can lead to more accurate regression results.

## Acknowledgments

J. Zhang, Z. Zhang, X. Xiao, M. Winslett, and Y. Yang are supported by the Agency for Science, Technology, and Research (Singapore) under SERC Grant 102-158-0074. J. Zhang and X. Xiao are also supported by the Nanyang Technological University under SUG Grant M58020016 and AcRF Tier 1 Grant RG 35/09.